\documentclass[runningheads]{llncs}
\usepackage[T1]{fontenc}
\usepackage[dvipdfmx]{graphicx} %normally used settings
\usepackage{enumitem}
\usepackage{amsmath}%
\usepackage{mathtools}%
\usepackage{wasysym}%
\usepackage[ruled,vlined]{algorithm2e}
\usepackage{bm}
\usepackage{amssymb}
\LinesNumbered

\usepackage [pagewise]{lineno} 
\usepackage[final,mode=multiuser]{fixme}
\fxuseenvlayout{colorsig}
\fxusetargetlayout{color}
\FXRegisterAuthor{ll}{all}{LL}
\FXRegisterAuthor{si}{asi}{SI}
\FXRegisterAuthor{hb}{ahb}{HB}
\FXRegisterAuthor{tm}{atm}{TM}
\fxsetface{env}{}

\usepackage[sort,nocompress]{cite}
\graphicspath{{graphics/}}

\newcommand{\excludeLeft}{\mathit{excludeLeft}}
\newcommand{\excludeRight}{\mathit{excludeRight}}

\newcommand{\pList}{\mathit{pList}}
\newcommand{\start}{\mathit{start}}

\newcommand{\siarr}{\mathit{SI}}
\newcommand{\piarr}{\mathit{PI}}
\newcommand{\plarr}{\mathit{PL}}

\newcommand{\pcon}{\mathit{P_{concat}}}
\newcommand{\pscon}{\mathit{PS_{concat}}}
\newcommand{\pseq}{\mathit{P_{seq}}}

\newcommand{\SAP}{\mathit{SAP}}
\newcommand{\SAPS}{\mathit{SAPS}}

\newcommand{\nul}{\mathit{null}}
\newcommand{\lrs}{\mathit{lrs}}
\newcommand{\ans}{\mathit{ans}}
\newcommand{\pArray}{\mathit{pArray}}
\newcommand{\LCP}{\mathit{LCP}}
\newcommand{\SA}{\mathit{SA}}

\newcommand{\res}{\mathit{res}}

\newcommand{\successorOffset}{\mathit{successorOffset}}
\newcommand{\sPCount}{\mathit{sPCount}}
\newcommand{\occ}{\mathit{occ}}

\newcommand{\mtt}{\mathtt}

\usepackage{tabularx}

\begin{document}
%
%\title{Online algorithms for substrings matching length range, prefixes, and suffixes}
\title{Online algorithms for finding distinct substrings with length and multiple prefix and suffix conditions}
\titlerunning{Online algorithms for distinct substrings with length, prefix, suffix conditions}
\author{Laurentius Leonard\inst{1}\orcidID{0000-0001-8477-7033} \and \\
  Shunsuke Inenaga\inst{2}\orcidID{0000-0002-1833-010X} \and \\
  Hideo Bannai\inst{3}\orcidID{0000-0002-6856-5185} \and \\
  Takuya Mieno\inst{4}\orcidID{0000-0003-2922-9434}
}

\authorrunning{L. Leonard, S. Inenaga., H. Bannai, T. Mieno}
\institute{Department of Information Science and Technology, Kyushu University, Japan
  \email{laurentius.leonard.705@s.kyushu-u.ac.jp}\\
  \and
  Department of Informatics, Kyushu University, Japan
  \email{inenaga@inf.kyushu-u.ac.jp}\\
  \and
  M\&D Data Science Center, Tokyo Medical and Dental University, Japan
  \email{hdbn.dsc@tmd.ac.jp}\\
  \and
  Department of Computer and Network Engineering, University of Electro-Communications, Japan
  \email{tmieno@uec.ac.jp}}

%\email{\{inenaga,yuto.nakashima\}@inf.kyushu-u.ac.jp}}

%\url{http://www.springer.com/gp/computer-science/lncs} 

%
\maketitle              % typeset the header of the contribution
\begin{abstract}
  Let two static sequences of strings $P$ and $S$, representing prefix and suffix conditions respectively, be given as input for preprocessing. For the query, let two positive integers $k_1$ and $k_2$ be given, as well as a string $T$ given in an online manner, such that $T_i$ represents the length-$i$ prefix of $T$ for $1 \leq i \leq |T|$.
  In this paper we are interested in computing
  the set $\mathit{ans_i}$ of distinct substrings $w$ of $T_i$
  such that $k_1 \leq |w| \leq k_2$,
  and $w$ contains some $p \in P$ as a prefix and some $s \in S$ as a suffix. 
  More specifically, the counting problem is to output $|\mathit{ans_i}|$, whereas the reporting problem is to output all elements of $\mathit{ans_i}$, for each iteration $i$. %NEW
  Let $\sigma$ denote the alphabet size,
  and for a sequence of strings $A$, $\Vert A\Vert=\sum_{u\in A}|u|$.
  Then, we show that after $O((\Vert P\Vert +\Vert S\Vert)\log\sigma)$-time preprocessing,
  the solutions for the counting and reporting problems for each iteration up to $i$ can be output in $O(|T_i| \log\sigma)$ and $O(|T_i| \log\sigma + |\mathit{ans_i}|)$ total time.
  %cumulative time, respectively, where cumulative time refers to the total amount of running time up to iteration $i$, as opposed to the running time of only iteration $i$. %NEW_END
  %Then,
  %we show that after $O((\Vert P\Vert +\Vert S\Vert)\log\sigma)$-time preprocessing,		
  %the counting problem of outputting $|\mathit{ans_i}|$ on each iteration $i$ can be solved in $O(|T_i| \log\sigma)$ cumulative time,
  %while the reporting problem can be solved in $O(|T_i| \log\sigma + |\mathit{ans_i}|)$ cumulative time, with both problems requiring only $O(|T_i|+\Vert P\Vert + \Vert S\Vert)$ working space.
  The preprocessing time can be reduced to $O(\Vert P\Vert +\Vert S\Vert)$ for integer alphabets of size polynomial with regard to $\Vert P\Vert +\Vert S\Vert$.
  Our algorithms have possible applications to network traffic classification.

\end{abstract}
%
%
%\cite{LLMultiPSingleSPreprint}
\section{Introduction}
Pattern matching has long been a central topic in the field of string algorithms~\cite{TextAlgorithms},
leading to various applications, including
DNA analysis in bioinformatics~\cite{DanGusfield,GenomeScale}
as well as
packet classification~\cite{Choi2008,Fuchino2019,Togaonkar2004,RunBasedTrie} and anti-spam email filtering in network security~\cite{Pampapathi2006}. %former line break
%Suffix trees, along with suffix arrays and DAWGs, are specialized data structures that solve pattern matching problems in a straightforward manner while offering more flexibility, such as the possibility to be made to handle multiple strings\cite{FeiShi,Takagi2019,Inenaga2020Pointer}.

In this paper, we propose algorithms for counting and reporting distinct substrings of an online text $T$ that have some $p \in P$ as a prefix and some $s \in S$ as a suffix, and whose length is within the interval $[k_1..k_2]$, where $P$ and $S$ are static sequences of strings given as input for preprocessing, and integers $k_1, k_2$ and the characters of $T$ are given as query.
A similar yet different problem where patterns are given in the form of a pair of a prefix and a suffix condition, i.e. $p\Sigma^*s$ patterns, rather than a pair of sequences where one is of prefixes and the other of suffixes,
is well-studied as the \emph{followed-by} problem~\cite{Baeza1996,Manber1991} or the \emph{Dictionary Recognition with One Gap (DROG)}~\cite{Amir2020,Levy2020,Shalom2021} problem.
The problem of this paper is also of importance
with possible applications in network traffic classification: All the application signatures in~\cite{Sen2004}, for example, can be expressed as an instance of our problem, as discussed in Section~\ref{SecSignatures} and demonstrated in Appendix~\ref{AppendixInputSet}.
Note that while \sinote*{which method is this?}{} traffic classification via these signatures was shown to be highly accurate, there are still cases of false positives. When analyzing which patterns give false positives, we may be interested in which patterns match the signatures, in which case the distinct condition of our problem helps prevent wasting computation time on repeated occurrences of each pattern.
%Another possible application is in checking whether a word $w$ with $|w|\geq |k|$ belongs to a $k$-$TSS$ language or a $(k,r)$-$TTSS$ language
\hbnote*{This doesn't contain much useful information since it requires knowledge of reference.
  Perhaps the idea of k-TSS, (k,r)-TTSS language needs to be described.
}{%
}%
Another possible application is in computing the distinct substrings of an online text $T$ whose lengths are at least $k$ and belong to a $(k,r)$-$\textit{TTSS}$ language~\cite{Ruiz1998};
the words of length $\geq k$ in a $(k,r)$-$\textit{TTSS}$ language defined by the 4-tuple $(I_k, F_k, T_{k,r}, g)$
are the words that have some element of $I_k$ as a prefix, some element of $F_k$ as a suffix, and includes, for each $t\in T_{k,r}$, at most $g(t)$ occurrences of $t$ as substring.
Here, the elements of $I_k$ and $F_k$ are strings of length $k-1$ and the elements of $T_{k,r}$ are strings of length between 1 to $k$ inclusive, and $g$ is a function that projects $T_{k,r}\rightarrow \{0,1,\cdots ,r-1\}$.
A direct application of our algorithms can check whether $w$ fulfills the prefix and suffix condition, while the condition of restricted segments, i.e. the number of occurrences of $T_{k,r}$ can also be considered by implementing the following modification:
for each iteration $i$, maintain the minimum start-index $\res$ of the suffix of $T_i$ that meets the condition of restricted segments, and use it to exclude any suffixes longer than $T_i[\res ..i]$ from the solution.
% by keeping count of such suffixes among those corresponded to by elements of $\pList$ introduced in this paper.
%ExcludeRes->Excludes from left (the longest suffixes have the most occurrences of restricted segments as substrings, and thus most likely to be excluded)
%can take the no. of occurrences of t as time complexity, if we use AC automaton to detect t occurrences?

Our proposed algorithms take
$O((\Vert P\Vert + \Vert S\Vert)\log\sigma)$ preprocessing time,
while processing $T$ itself in an online manner and outputting the solutions up to iteration $i$ takes
$O(|T_i| \log\sigma)$ and
$O(|T_i| \log\sigma + |\ans_i|)$ cumulative time
for the counting and reporting problems respectively, using $O(|T_i|+\Vert P\Vert+\Vert S\Vert)$ working space.
Here, $T_i$ denotes the length-$i$ prefix of $T$,
$\Vert P\Vert$ and $\Vert S\Vert$ denote the total length of strings in $P$ and $S$ respectively, $\sigma$ is the alphabet size, $|\ans_i|$ is the number of substrings reported for each $T_i$,
and cumulative time refers to the total amount of running time up to iteration $i$, as opposed to the running time of only iteration $i$.
In addition, the preprocessing time can be reduced to $O(\Vert P\Vert + \Vert S\Vert)$ in the case of integer alphabets of size polynomial in $\Vert P\Vert + \Vert S\Vert$.

\hbnote*{I didn't understand after "exactly that ...".
  Also, can't we say it is a generalization, rather than just "differs"?}{%
  Also note that, the problems addressed in this paper differ from those of~\cite{LLMultiPSingleSPreprint}, in which a different set of solution strings is output for each $p \in P$ where each solution must have that specific element of $P$ as a prefix, unlike the problem in this paper where only a single set of solution strings is output, where its elements can have any $p\in P$ as a prefix.
}\llnote{I'm not sure yet about whether we can say it is a generalization considering the differences that the two problems have, but I have changed the description of the referenced problem to hopefully make it clearer.}
Also, in~\cite{LLMultiPSingleSPreprint} there is only one suffix condition and no length condition, and the algorithm is offline w.r.t. $T$, unlike in this paper.

\section{Preliminaries and definitions}

\subsection{Strings}
\label{subsecPrelim}
Let $\Sigma$ be an alphabet of size $\sigma$. An element of the set $\Sigma^*$ is a string.
\sinote*{added}{%
The length of a string $w$ is denoted by $|w|$.
The empty string is denoted by $\varepsilon$. That is, $|\varepsilon|=0$.
For a string $w = pts$, $p$, $t$, and $s$ are called a
\emph{prefix}, \emph{substring}, and \emph{suffix} of $w$, respectively.
A prefix $p$ (resp. suffix $s$) of a string $w$ is called a
\emph{proper prefix} (resp. \emph{proper suffix}) of $w$
if $|p| < |w|$ (resp. $|s| < |w|$).
}
For a string $w$,
$w[i]$ denotes the $i$-th symbol of $w$ for $1 \leq i \leq |w|$, and
$w[i..j]$ denotes the substring $w[i]w[i+1]\cdots w[j]$ for $1 \leq i \leq j \leq |w|$.
For a sequence $S$ of strings, let $\Vert S\Vert = \sum_{u\in S}|u|$.

\subsection{Suffix array and LCP array}
The suffix array~\cite{Manber1993} of a string $w$ is a lexicographically sorted array of suffixes of $w$, where each suffix is represented by its start-index.
The LCP array is an auxiliary array commonly used alongside the suffix array, that stores the length of the longest common prefix of each adjacent pair of suffixes in the suffix array.
More specifically, if $\SA$ and $\LCP$ are the suffix array and LCP array of the same string $w$, for $x \in [2..|w|]$, $\LCP[x]$ is the length of longest common prefix of the suffixes $w[\SA[x]..|w|]$ and $w[\SA[x-1]..|w|]$.
In this paper, we will use suffix arrays for some strings, % one by one,
in which we denote by $\LCP$ the LCP array of the same string of the suffix array being discussed.
It is well-known that the suffix array and LCP array of a string $w$ can be built in $O(|w|)$ time for integer alphabets of polynomial size in $|w|$~\cite{Karkkainen2006, Kasai2001, Kim2005}, and in $O(|w| \log \sigma)$ time for general ordered alphabets~\cite{Ukkonen}.

\subsection{The problems}
%Restating the problems in Section \ref{secIntro} using the notations of Section \ref{subsecPrelim}, the problems considered in this papers are formally written as follows:
The problems considered in the paper are as follows.

\begin{definition}[Online substring counting and reporting problem with distinctness, multiple prefixes, multiple suffixes and length range conditions]
  \label{DefProblem}
  Given two sequences of strings $P=(p_1,\cdots , p_n)$ and $S=(s_1,\cdots ,s_m)$, two integers $k_1$ and $k_2$, and a string $T$ given in an online manner
  \hbnote*{slightly reworded}{%
    (i.e.,
    $T_0 = \varepsilon$ and for each iteration $i = 1,\dots, |T|$,
    the $i$-th character is appended to $T_{i-1}$ to form $T_i$),
  }%
  let $\ans_i$ denote the set of distinct substrings of $T_i$ that have some $p \in P$ as a prefix, some $s \in S$ as a suffix, and whose length falls within the interval $[k_1..k_2]$.
  %Formally, 
  %\begin{equation}
  %ans_i = \{x \in \Sigma^k \mid k \in [k_1, k_2], x \in Sub(T_i), Pre(x) \cap P \neq \emptyset, Suf(x) \cap S \neq \emptyset\} %[PRESUFSUB]
  %\end{equation}

  \begin{itemize}
    \item[] \textbf{The counting problem.} On each iteration $i$, output $|\ans_i|$.
    \item[] \textbf{The reporting problem.} On each iteration $i$, output $\ans_i \setminus \ans_{i-1}$.
  \end{itemize}
\end{definition}

\hbnote*{do we need this?}{%
  This paper excludes the empty string $\varepsilon$ from the solutions.
}%

\section{Algorithm}
%rough sketch
\subsection{Sketch of algorithm}
\label{secsketch}
%define new
%add figure
In this section, we describe the general idea of our algorithm.
During each iteration $i$, we need either to compute the size of $\ans_i\setminus \ans_{i-1}$ to add it to the counting solution, or to report all its elements.
All elements of $\ans_i\setminus \ans_{i-1}$ must be suffixes of $T_i$, and thus for the suffix condition,
$T_i$ itself must have some element of $S$ as a suffix;
otherwise clearly $\ans_i\setminus \ans_{i-1} = \emptyset$ and there is no need to output a solution for the current iteration.
Thus, let us consider the case where $T_i$ has at least one element of $S$ as a suffix, and call the shortest of them $s$. %As any element of $\ans_i\setminus \ans_{i-1}$ needs to occur in $T_i$ but not in $T_{i-1}$, it suffices to look at the suffixes of $T_i$.
To help keep track of which suffixes of $T_i$ fulfill the prefix condition,
let us maintain a linked list $\pList$ that contains
in increasing order, all distinct indices $j$ such that there is an element of $P$ that occurs in $T_i$ with start-index $j$.
Clearly, the elements of $\pList$
%\hbnote*{usage may be a bit informal}{%
%  bijects to
%}%
\sinote*{modified}{%
  represent a bijection to the
}%
suffixes of $T_i$ that have an element of $P$ as a prefix, which are candidates for elements of the solution set $\ans_i\setminus \ans_{i-1}$.
Specifically, for all $j\in \pList$, the string $u = T_i[j..i] \in \ans_i\setminus \ans_{i-1}$ iff $u$ fulfills all the following conditions:
\begin{enumerate}[label=(\alph*)]
  \item $u$ has $s$ as a suffix.
  \item $u$ does not occur in $T_{i-1}$.
  \item $|u| \geq k_1$ .
  \item $|u| \leq k_2$.
\end{enumerate}

\begin{figure}[h!]
  \centering
  \includegraphics[width=12.1cm]{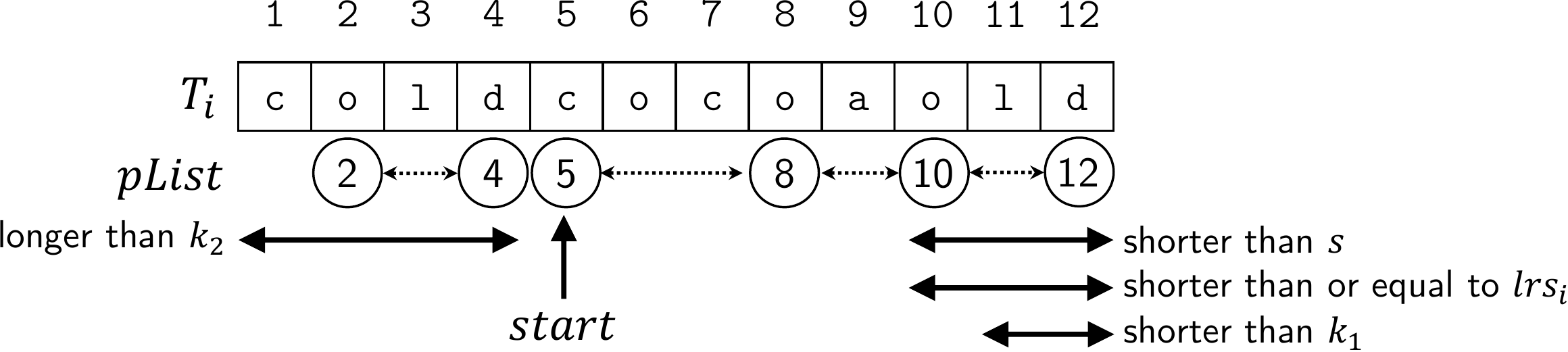}
  \caption{A visualized example of $\pList$.}
  \label{fig:pList}
\end{figure}

Here, (a) $u$ has $s$ as a suffix iff $|u| \geq |s|$, and (b) $u$ does not occur in $T_{i-1}$ iff $|u| > |\lrs_i|$, where $\lrs_i$ denotes the longest repeating suffix of $T_i$, i.e. the longest suffix of $T_i$ that occurs at least twice in $T_i$. Thus, conditions (a) to (c)
\hbnote*{reworded}{%
  set a lower bound for the length of suffixes of $T_i$ corresponding to elements of
  $\pList$ that can be a solution
}%
%excludes
%\hbnote*{couldn't parse}{%
%  some of the shortest suffixes of $T_i$ corresponded to by elements of $\mathit{pList}$ from the solution,
%}%
while condition (d)
\hbnote*{reworded}{%
  sets an upper bound.  %excludes some of the longest suffixes.
}
If we visualize $\pList$ horizontally as shown in Figure \ref{fig:pList}, conditions (a) to (c) exclude some elements from the right while condition (d) excludes some element from the left.

%%% insert plist figure

Take the maximum among the number of elements excluded by conditions (a) to (c) and denote it by $\excludeRight$, and denote the number of elements excluded by condition (d) by $\excludeLeft$.
Then, $|\ans_i\setminus \ans_{i-1}|$
$=$
$\max{}(0, |\pList |-\excludeLeft-\excludeRight)$, giving us the solution for the counting problem. %minmaxnotation %enclosing the whole expression in max's {} makes it overflow, hence the \max{}
For the reporting problem, let us maintain $\start$, a pointer to the smallest element in $\pList$ not excluded by condition (d). Then report all elements traversed by starting at $\start$ and moving to the right $|\ans_i\setminus \ans_{i-1}|-1$ times.

\begin{example}
  \label{exPList}
  The example in Figure~\ref{fig:pList} occurs when $T_i=\mtt{coldcocoaold}$, $P=(\mtt{cave}, \mtt{coco}, \mtt{cocoa}, \mtt{d}, \mtt{oao}, \mtt{old})$,
  $S=(\mtt{aold}, \mtt{oaold})$, $k_1=3$, and $k_2=8$.
  Then, $\pList=(2,4,5,8,10,12)$. $10, 12$ correspond to $\mtt{old}, \mtt{d}$ which are shorter than $s=\mtt{aold}$, and thus excluded by condition (a). $\lrs_i=\mtt{old}$ and thus condition (b) also excludes $10$ and $12$, while condition (c) excludes only $12$ which corresponds to $\mtt{d}$.
  Therefore, $\excludeRight=\max{(2,2,1)}=2$. Meanwhile, condition (d) excludes $2$ and $4$ which correspond to $\mtt{oldcocoaold}$ and $\mtt{dcocoaold}$ and so $\excludeLeft=2$.%minmaxnotation
  Thus, we have that $|\ans_i\setminus \ans_{i-1}|=\max{(0, 6-2-2)}=2$. For the reporting solution, we have that $\start$ points to $5$. Traversing $2$ elements starting from $8$ gives us $5$ and $8$, each corresponding to $\mtt{cocoaold}$ and $\mtt{oaold}$, exactly the elements of $\ans_i\setminus \ans_{i-1}$.%minmaxnotation
\end{example}

\subsection{Removing redundant elements}

We say that $p_k\in P$ is \emph{redundant} iff there exists $p_{k'}$ of $P$ s.t. either $p_k$ has $p_{k'}$ as a proper prefix, or $p_k = p_{k'} \wedge k>k'$.
Similarly, $s_k\in S$ is \emph{redundant} iff there exists $s_{k'}$ of $S$ s.t. either $s_k$ has $s_{k'}$ as a proper suffix, or $s_k = s_{k'} \wedge k>k'$.

It is not hard to see why they are called redundant; when
multiple copies of the same string exist in $P$, keeping only one copy suffices, and when
$p_k\in P$ has $p_{k'}\in P$ as a proper prefix, the strings that have $p_k$ as a prefix is a subset of strings that have $p_{k'}$ as a prefix,
and thus the solution remains the same even if we delete $p_k$ from $P$. The same can be said for redundant elements of $S$.

As one part of the preprocessing, we rebuild $P$ and $S$ so that the redundant elements are deleted.
First, we describe how to rebuild $P$.
Let $\pseq=\$p_1\$p_2\$\cdots \$p_n\$$, where $\$ \not \in \Sigma$ and $\$ \prec c$ for all $c\in \Sigma$,
and let $\SA$ be the suffix array of $\pseq$.

Then, each $\SA[x]$ for $x\in [2..n+1]$ corresponds to the start-index of $\$p$ in $\pseq$, for some $p\in P$. %change: using n, m for size of P and S
Starting from $x=2$,
output the corresponding $p$, namely the
unique $p\in P$ s.t. $\$p\$$ occurs on index $\SA[x]$.
Then, increment $x$ (at least once) until we have that $\LCP[x] < |\$p|$, i.e. until we find an index that corresponds to $\$p'$ where $p' \in P$ does not have $p$ as a prefix.
Output the element of $P$ corresponding to the new $x$, then again increment $x$ in the same manner.
Repeat this until $x>n+1$, and we have that all non-redundant elements of $P$ are output. %change: using n, m for size of P and S

\hbnote*{need mention of integer alphabet? - perhaps write about SA build time in preliminaries}{%
  Other than the construction of the suffix array and LCP array, clearly this takes $O(\Vert P\Vert)$ time, and the same method can be used to compute non-redundant elements of $S$:
}%
Let $S^{-1}=(s_1^{-1},\ldots ,s_m^{-1})$ be the sequence of reversed elements of $S$, then apply the above algorithm to $S^{-1}$ and reverse each string in the output to get the non-redundant elements of $S$.
Thus, both $P$ and $S$ are rebuilt to exclude redundant elements in $O(\Vert P\Vert + \Vert S\Vert)$ time, in addition to the construction time of the suffix array and LCP array, which depends on the alphabet.

\begin{example}
  Let $P=(\mtt{abc},\mtt{ab},\mtt{acc},\mtt{ab},\mtt{cab})$. Then,
  $\pseq=\mtt{\$abc\$ab\$acc\$ab\$cab\$}$ and we have the table as shown in Figure~\ref{fig:reduntable}.
  $x=2$ corresponds to the occurrence of $\mtt{\$ab}$, which occurs on $\pseq$ on index $\SA[2]=5$.
  Thus, $\mtt{ab}$ is determined to be non-redundant, and we have that $|\mtt{\$ab}|=3$, so increment $x$ until
  we have that $\LCP[x] < 3$. This skips over $x=3,4$, correctly determining their corresponding elements of $P$, namely the second $\mtt{ab}$ and $\mtt{abc}$ to be redundant.
  When $x=5$, $\LCP[x] = 2 < 3$ and so the corresponding $p = \mtt{acc}$ is output. Similarly, for $x=6$, $\LCP[x] = 1 < 4$ and thus $\mtt{cab}$ is output.
  Afterwards, $x$ is incremented beyond the interval $[2..n+1]$ and thus the algorithm terminates and the non-redundant elements $(\mtt{ab}, \mtt{acc}, \mtt{cab})$ are output. %change: using n, m for size of P and S

  \begin{figure}[h!]
    \centering
    \includegraphics[width=9cm]{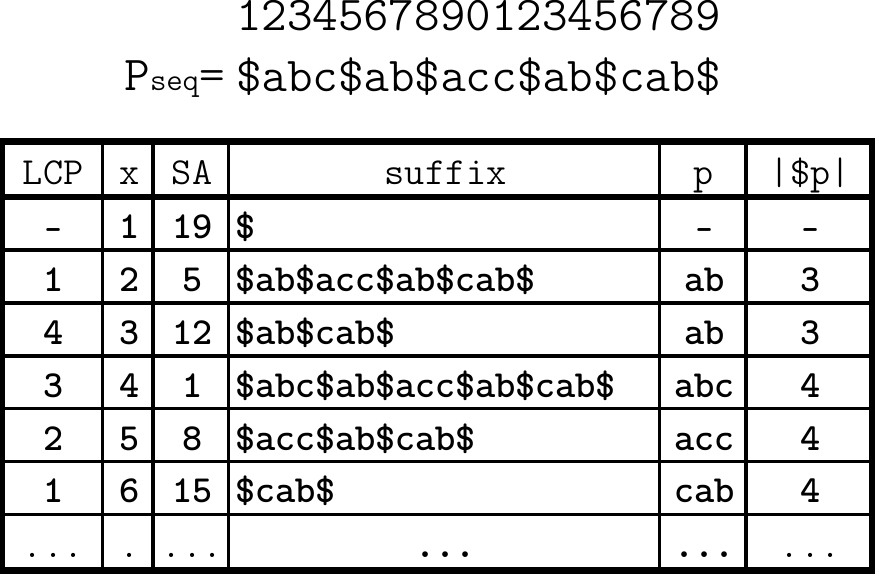}
    \caption{The table for $\pseq=\$abc\$ab\$acc\$ab\$cab$}
    \label{fig:reduntable}
  \end{figure}

\end{example}

For the rest of the paper, we will assume that the above preprocessing is done and thus \sinote*{maybe use $\langle p_1, \ldots, p_n \rangle$ as $P$ and $S$ are now sequences?}{$P=(p_1, \ldots ,p_n)$ and $S=(s_1, \ldots, s_m)$} from this point refer to the rebuilt sequences that have no redundant elements.
\llnote*{Changed to use ( ) instead of $\{ \}$ to match with the rest of paper. (can also use $\langle \rangle$ if preferred)}

\subsection{Detecting $P$ and $S$ occurrences}
As discussed, on each iteration $i$ we need to detect whenever an element of $S$ occurs as a suffix of $T_i$.
Additionally, we will also need to detect when an element of $P$ occurs as a suffix of $T_i$, in order to maintain $\pList$.
This can be done by building an Aho-Corasick automaton for $P$ and $S$ separately.
Constructing both automata takes $O((\Vert P\Vert + \Vert S\Vert) \log \sigma)$ preprocessing time in general,
and $O(\Vert P\Vert + \Vert S\Vert)$ time in the case of integer alphabets of size polynomial with regard to $\Vert P\Vert + \Vert S\Vert$~\cite{Dori2005}. %CONSIDER: This can be moved into summary or preliminaries, if needed
Running each automaton up to iteration $i$ takes $O(|T_i|\log \sigma + \occ)$ cumulative time, where $\occ$ is the number of occurrences detected.
Here, the occurrences of elements of $P$ in $T_i$ must have distinct start-indices, as two occurrences with a shared start-index imply that one of them is redundant.
Similarly, occurrences of elements of $S$ must have distinct end-indices and thus $\occ \in O(|T_i|)$ for both automata, and so the cumulative running time becomes $O(|T_i|\log \sigma)$.

\subsection{Maintaining $\pList$}
To maintain $\pList$,
whenever some $p\in P$ is detected to occur as a suffix of $T_i$, its start-index $j$ needs to be added to $\pList$ while maintaining the increasing order.
Doing this naively would take $O(|\pList|) = O(|T_i|)$ time for every insertion which gives quadratic time overall, so a more efficient scheme is necessary.

During any iteration $i$, the elements of $P$ that occur as suffixes of $T_i$ are detected in decreasing order of length, because we use Aho-Corasick automaton.
Let $p$ be such an element detected, and $j$ be the start-index of its occurrence, i.e. $j=i-|p|+1$.
We need to add $j$ into $\pList$ so that the increasing order of its elements are maintained.
%
%such that the increasing order holds.

To do that, we need to find the minimum $j'$ among the current elements of $\pList$ such that $j' > j$.
Here, $j'$ being an element of $\pList$ implies that $j'$ corresponds to an occurrence of $p'\in P$ starting at $j'$ and ending at some $i'\leq i$.
We can see that in fact $i'< i$, for if $i'=i$,
$j'$ was added to $\pList$ in the current iteration $i$ before $j$, while $j'>j \wedge i'=i$ implies $|p'|<|p|$, contradicting the fact that the Aho-Corasick automaton detects the occurrences in decreasing order of length.
Thus, $j'>j$ and $i'<i$, meaning the occurrence of $p'$ falls completely within $p[2..|p|-1]$.

Our scheme is then as follows: %Let $P=(p_1, \cdots p_n)$. Then 
We precompute, for each $p_k\in P$, the minimum value $y$ such that there is some $p_{k'} \in P$ that occurs in $p_k[2..|p_k|-1]$ on start-index $y+1$. If there is no such $p_{k'}$, then let $y=\infty$.
Then, we will store the values on the array $\successorOffset$ that maps each $p_k\in P$ to its corresponding $y$.

Additionally, maintain also an array $\pArray$ such that $\pArray[j]$ points to the element of $\pList$ whose value is $j$ if it exists, or $\nul$ otherwise.
Then, whenever some $p_k\in P$ occurs as a suffix of $T_i$ with start-index $j$,
we can just add $j$ into $\pList$ exactly before the element pointed to by $pArray[j+\successorOffset[k]]$.
Clearly, once $\successorOffset$ is computed, adding each element of $\pList$ takes only constant time and thus maintaining $\pList$ and $\pArray$ takes cumulative $O(|T_i|)$ time, as the number of elements of $\pList$ for any given iteration $i$ is bounded by $|T_i|$.

\begin{figure}[h!]
  \centering
  \includegraphics[width=12cm]{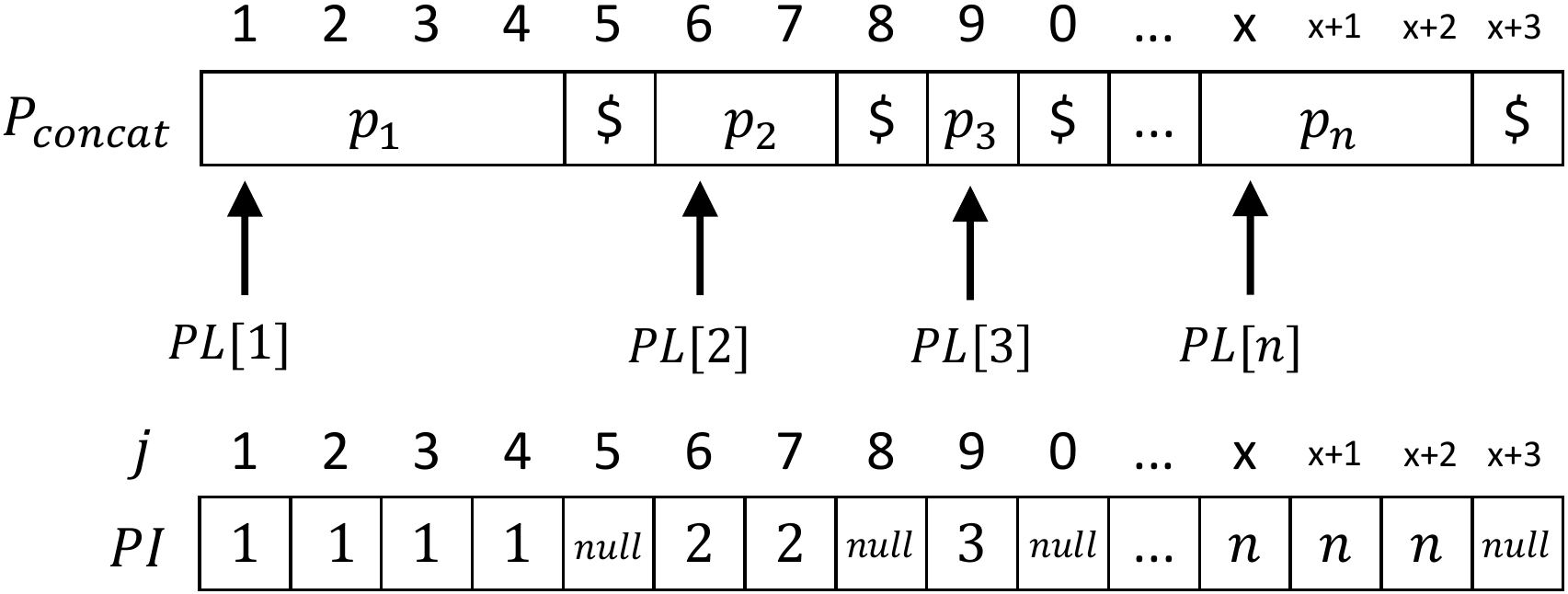}
  \caption{Example of $\plarr$ and $\piarr$.}
  \label{fig:pij}
\end{figure}

\paragraph{Computing $\successorOffset$.}
Let $\pcon = p_1\$p_2\$\cdots \$p_n\$$.
Note that it differs from $\pseq$ not only with regard to the positioning of $\$$, but also in that redundant elements of $P$ are not included.
For $k \in [1..n]$,
let $\plarr[k]$ denote the start-index of $p_k$ in $\pcon$, i.e.
$\plarr[k]=1+\sum_{k'\in [1..k)}{(|p_{k'}|+1)}$.
For $j \in [1..|\pcon|]$,
let $\piarr[j]$ be the index of the element of $P$ covering index $j$ in $\pcon$. %excluding when $P_{concat}[j]=\$$ or $j$ is the end-index of said element of $P$, in which case $PI[j]=null$.
That is, $\piarr[j]=k$ where $j\in[\plarr[k]..\plarr[k]+|p_k|-1]$ if such $k$ exists, otherwise $\piarr[j]=\nul$. %開始位置、終了位置をここで除くような定義すると後でその条件をチェックしなくよくなるが、exclusion by sのところに再利用できなくなります
See Figure~\ref{fig:pij} for an example.
%Formally,
%\begin{align}
%PI[j] &=
%\begin{dcases}
%k  \textbf{\,\,if } \exists k\in [1,n] \textbf{\,\, s.t.\,\,} j \in [PL[p_k], PL[p_k]+|p_k|-2]\\
%null \textbf{\,\, otherwise.}
%\end{dcases} \\
%\end{align}

Next, we construct the suffix array $\SAP$ of $\pcon$.
%maybe cite linear time construction of suffix arrays?
We can then compute $\successorOffset$.
The general idea is that for each $p_k$, we find all its occurrences using the suffix array, and when the occurrence falls within some $p_{k'} \in P$, we update $\successorOffset[k']$.
A more detailed description is as follows:

\begin{itemize}
  \item Initialize $\successorOffset[k]=\infty$ for all $p_k\in P$ .
  \item
        Using the LCP array,
        find the subinterval $[\ell..r]$ in $\SAP$ that corresponds to occurrences of $p_k$, i.e. each of $\SAP[x]$ for all $x \in [\ell..r]$ corresponds to some start-index of occurrences of $p_k$ in $\pcon$.

  \item For each such occurrence, whenever it falls within $p_{k'}[2..|p_{k'}|-1]$ for some $p_{k'}\in P$, then \sinote*{what do you mean by this ``min''?}{}assign to $\successorOffset[k']$ the minimum value between itself and the offset distance $\SAP[x]-\plarr[k']$.
        Formally, for all $x\in [\ell..r]$, if
        $\pcon[\SAP[x]+|p|]\neq \$$, 
%        $\piarr[\SAP[x]]\neq \, \nul \, \wedge \, \pcon[\SAP[x]+|p|]\neq \$$,  %It used to be that for occurrences of p_k in P_seq, the first and last (before $) index has its PI value null. We don't do this anymore now, so the only places where PI is null, are the $'s. SAP[x] cannot be on a $ occurrence, since SAP[x] bijects to start-indices of occurrences of p_k.
%        $\piarr[\SAP[x]]\neq \, \nul \, \wedge \plarr[\piarr[\SAP[x]]] \neq \SAP[x] \wedge P_{concat}[\SAP[x]+|p|]\neq \$$,   %second condition removed. It excludes occurrences pf p_k as prefix of p_{k'}, which only happens when p_k=p_{k'} as otherwise p_k is redundant. The case where p_k=p_{k'} is already excluded by the third condition, which excludes occurrences as suffix of p_{k'}.
        assign to $\successorOffset[\piarr[\SAP[x]]]$
        the following value:
        \begin{equation}
          \min{(\successorOffset[\piarr[\SAP[x]]],\SAP[x]-\plarr[\piarr[\SAP[x]]])} %minmaxnotation
        \end{equation}
\end{itemize}

Computing $\pcon$ and $\piarr$ trivially takes $O(\Vert P\Vert)$ time. %Constructing $\SAP$ takes $O(|P_{concat}|) = O(\Vert P\Vert)$ time.
To compute the subinterval corresponding to occurrences of $p_k$,
simply find the longest subinterval $[\ell..r]$ of $\SAP$ that includes the index $\SAP^{-1}[\plarr[k]]$ and $\LCP[x]\geq |p_k|$ for all $x \in [\ell+1..r]$. 
This takes linear time w.r.t. to the subinterval length, which all adds up to the total number of occurrences of elements of $P$ in $\pcon$. As no occurrence may share a start-index, it is bounded by $|\pcon|\in O(\Vert P\Vert)$.
Thus, this preprocessing takes $O(\Vert P\Vert)$ time, in addition to the time required to compute $\SAP$ and $\LCP$ which depends on the alphabet.
%Thus, this preprocessing can be done in $O(\Vert P\Vert \log \sigma)$ time.

\subsection{Computing $\excludeLeft$, $\excludeRight$, and $\start$}
As discussed, the algorithm runs the Aho-Corasick automaton for $S$
to check whether there is some $s\in S$ that occurs as a suffix of $T_i$, for each iteration $i$. %This is done using Aho-Corasick automaton, taking
\sinote*{Is this $O(\Vert S \Vert)$ for integer alphabet?}{}%  $O(\Vert S\Vert \log \sigma)$}%%construction time and cumulative $O(|T_i|+\log \sigma)$ running time, similar to the detection of $P$ occurrences.$O(|T_i|+\log \sigma)$ cumulative running time, similar to the detection of $P$ occurrences.
\llnote*{Since the method and time complexity for detecting elements of $S$ were described in Section 3.3, I have opted to omit this part.}{}
In case such $s$ exists, all of the computations below are performed, otherwise only maintaining $\start$ is necessary.

\subsubsection{$\excludeRight$}

\paragraph{Exclusion by $s$.}
When $s\in S$ occurs as a suffix of $T_i$,
the suffix $u=T_i[j..i]$ for each $j\in \pList$ is excluded from the solution
iff $j > i-|s|+1$.
That is, $u$ fails to meet condition (a) of Section~\ref{secsketch} iff
$j$ corresponds to an occurrence of $p\in P$ in $s[2..|s|]$.
Thus, if we preprocess the number of occurrences of
elements of $P$ that occur within $s[2..|s|]$ for each $s\in S$,
we can compute the number of elements of $\pList$ excluded by $s$ in constant time.

\begin{example}
  In the case shown by Example~\ref{exPList},
  the elements 10 and 12 are excluded by $s$, which we can obtain from the fact
  that there are two occurrences of elements of $P$ within $s[2..|s|]=\mtt{old}$; one each of $\mtt{old}$ and $\mtt{d}$.
\end{example}

A suffix array-based approach that is similar to what we used to compute $\successorOffset$ can be used here.
%Let $\pscon = p_1\$\cdots p_n\$s_1\$\cdots s_m$ for $P=\{p_1,\cdots p_n\}$ and $S=\{s_1[2..|s_1|],\cdots s_m[2..|s_m|]\}$,
Let $$\pscon = p_1\$\cdots p_n\$s_1[2..|s_1|]\$\cdots s_m[2..|s_m|],$$
and construct its suffix array $\SAPS$.
Define an array $\siarr$ such that $\siarr[j]=k$ when $j$ belongs to the part made up by $s_k[2..|s_k|]$ in $\pscon$, similar to $\piarr$.
Then, for all $p\in P$, compute the subinterval $[\ell..r]$ in $\SAPS$ corresponding to suffixes of $\pscon$ that start with $p$, again using $\SAPS^{-1}$ and LCP arrays.
For all $x \in [\ell..r]$, then increment $\sPCount[\siarr[\SAPS[x]]]$ by one, where $\sPCount$ is an array that maps each $s\in S$ to
\sinote*{something wrong?}{%
  the number of occurrences of elements of $P$ that occur within $s[2..|s|]$.
}\llnote*{Fix: the number of elements of $P$ occurrences that occur within $s[2..|s|]$.$\rightarrow$ the number of occurrences of elements of $P$ that occur within $s[2..|s|]$.}
Naturally, $\sPCount$ initially maps all elements of $S$ to zero before the counts are incremented.

The total of size of subintervals is bounded by the number of occurrences of elements of $P$ in $\pscon$, which is $O(|\pscon|) = O(\Vert P\Vert + \Vert S\Vert)$.
Thus, computing $\successorOffset$ takes $O(\Vert P\Vert + \Vert S\Vert)$ time, in addition to the construction time of $\SAPS$ and $\LCP$ which depends on the alphabet.
After the preprocessing, the number of elements of $\pList$ excluded by $s$ when $s$ occurs as a suffix of $T_i$ can be computed in constant time by simply referring to $\sPCount[s]$.

%for each s\in S, leftmost occurrence of p element in that s. (let's say it is leftP(s)) exclude when p=s.
%for each p\in P, number of p' \in P that occurs in p (let's say numP(p)) including itself
%Then, numP(leftP(s)) is excludedByS.

\paragraph{Exclusion by $\lrs_i$.}
%Ukkonen's algorithm
It is known that Ukkonen's algorithm\cite{Ukkonen} maintains the locus of $\lrs_i$ during each iteration $i$ where it builds the suffix tree of $T_i$, and thus we can compute its start-index $i-|\lrs_i|+1$ by simply running Ukkonen's algorithm.
Then, clearly $j\in \pList$ is excluded by condition (b) of Section~\ref{secsketch} iff $j \geq i-|\lrs_i|+1$.
Furthermore, the start-index is non-decreasing between iterations, i.e. $i-|\lrs_i|+1 \geq (i-1)-|\lrs_{i-1}|+1$ for any iteration $i$.
Thus, we can always maintain the count
$|\{j\in \pList \mid j \in [i-|\lrs_i|+1..i] \}|$ for each $i$ as follows:
%$excludedByLrs = |\{j\in \mathit{pList} \mid j \in [lrs_i..i] \}|$ for each $i$ as follows:

\begin{itemize}
  %\item During each iteration, consider the length one suffix $T_i[i]$; if $|lrs_i|\geq 1$ and $i\in pList$, then increment the count by one. %  \item During each iteration, $i$ is incremented by one, so check if there exists $i\in \pList$ and if so, increment the count by one.
  \item During some iterations, $i-|\lrs_i|+1 > (i-1)-|\lrs_{i-1}|+1$ which we will know from Ukkonen's algorithm. In that case, for each $j \in [(i-1)-|\lrs_{i-1}|+2.. i-|\lrs_i|+1]$ such that $j\in \pList$,
        decrement the count by one.
  \item Whenever a new element $j$ is added to $\pList$ such that $j\in i-|\lrs_i|+1$, increment the count.
\end{itemize}
We can check whether $j\in \pList$ in constant time for any $j$ using $\pArray$,
and both the left end $i-|\lrs_i|+1$ and right end $i$ of the interval only ever increases and is within $1$ to $i$,
so the above method takes cumulative $O(|T_i|)$ time, dominated by the runtime of Ukkonen's algorithm which is cumulative $O(|T_i| \log \sigma)$ time.
%mentioning this instead of only |T_i| log σ, because this complexity also appears in next paragraph.

\paragraph{Exclusion by $k_1$.}
The same approach can be used to find the number of elements of $\pList$ excluded by $k_1$:
we maintain the number of $j\in \pList$ such that
$i-j+1<k_1 \Leftrightarrow j\in [i-k_1+2..i]$ for each iteration $i$.
  Since both endpoints of this interval can only increase and are always between $1$ to $i$ inclusive,
  maintaining the count can be done in cumulative $O(|T_i|)$ time.

  \subsubsection{$\excludeLeft$ and $\start$}
  \paragraph{Maintaining $\excludeLeft$.}
  Similarly,
$\excludeLeft$
  is the number of $j\in \pList$ such that
$i-j+1>k_2 \Leftrightarrow j\in [1..i-k_2]$, and the same approach can be used to maintain this count in $O(|T_i|)$ time.

  \paragraph{Maintaining $\start$.}
  We can easily maintain $\start$ so that it points to the minimum element of $\pList$ of value at least $i+1-k_2$ in cumulative $O(|T_i|)$ time as follows:

  \begin{itemize}
    \item Initialize $\start$ to $\nul$.
    \item During each iteration $i$, if $\start$ is not $\nul$ and $\start$ < $i+1-k_2$, then let $\start$ point to the next element in $\pList$.
    \item Whenever a new element $j$ is added to $\pList$ such that $j \geq i+1-k_2$, if $\start = \nul$ or $\start > j$, let $\start$ point to $j$.
  \end{itemize}

  \subsection{Summarizing the algorithm}
  In the preprocessing, suffix arrays and LCP arrays are constructed and used to remove redundant elements of $P$ and $S$, as well as compute $\successorOffset$. Additionally, Aho-Corasick automata for $P$ and $S$ are built.
  For general ordered alphabets, this preprocessing takes $O((\Vert P\Vert + \Vert S\Vert)\log \sigma)$ time, with the construction time for Aho-Corasick automata and suffix arrays being the bottleneck.
  In the case of integer alphabets of size polynomial w.r.t. $\Vert P\Vert +\Vert S\Vert$,
  the construction times, and consequently the whole preprocessing time, can be reduced to $O(\Vert P\Vert + \Vert S\Vert)$ time.

  For the query processing time up to any iteration $i$,
$\excludeLeft$, $\excludeRight$, and $\start$ are computed in $O(|T_i|\log \sigma)$ cumulative time,
  giving us the solution for the counting problem.
  For the reporting problem, we traverse a total of $|\ans_i|$ elements in $\pList$, giving us $O(|T_i|\log \sigma + |\ans_i|)$ cumulative time, assuming the solution strings are output in the form of index pairs.

  Additionally, all the data structures used require only $O(|T_i| + \Vert P\Vert + \Vert S\Vert)$ total working space.
  Thus, we have the following results.

  \begin{theorem}
  \label{theor}
    There are algorithms that solve the counting and solving problems from Definition~\ref{DefProblem} for general ordered alphabets,
    such that after
    $O((\Vert P\Vert + \Vert S\Vert)\log \sigma)$ preprocessing time,
    the solutions are output for each iteration up to $i$
    in $O(|T_i|\log \sigma)$ time for the counting problem, and
    $O(|T_i|\log \sigma + |\ans_i|)$ time for the reporting problem,
    using $O(|T_i| + \Vert P\Vert + \Vert S\Vert)$ total working space.
%    The preprocessing time can be reduced to $O(\Vert P\Vert + \Vert S\Vert$) time % moved to corollary
%    in case of integer alphabets of size polynomial with regard to $\Vert P\Vert +\Vert S\Vert$.%
  \end{theorem}

\begin{corollary}
The preprocessing time in Theorem \ref{theor} can be reduced to $O(\Vert P\Vert + \Vert S\Vert)$ time in case of integer alphabets of size polynomial with regard to $\Vert P\Vert +\Vert S\Vert$.
\end{corollary}

  \section{Applying the algorithm for traffic classification}
  \label{SecSignatures}
%TODO REFER TO APPENDIX
We show in Appendix~\ref{AppendixInputSet} the input sets that match each of the application signatures described in~\cite{Sen2004}.

Note that the signatures shown in \cite{Sen2004} are generally characterized in the format of $p$ followed by $s$, where $p\in P$ and $s\in S$ for sets or lists $P$ and $S$.
In general, this differs from our problem in that occurrences of $p$ and $s$ overlapping should not be counted as a match. For example, $\mtt{ab}$ followed by $\mtt{bc}$ means $\mtt{abc}$ should not be counted
as a match, while our algorithms do count this as a match.
Nevertheless, such matches, which would be erroneous in the context of implementing the signatures, do not occur with the input sets listed in Appendix~\ref{AppendixInputSet}, as the elements of $P$ and $S$ simply cannot overlap. For example, with Gnutella signatures each element of $P$ ends with the character $\mtt{:}$, which no element of $S$ contains, so no string $w\in \Sigma^*$ exists such that has some $p\in P$ as prefix, $s\in S$ as suffix, and $p$ and $s$ overlap (i.e. $|w| < |p|+|s|$).
Note also that this problem also differs from the followed-by problem of~\cite{Baeza1996,Manber1991}, which does share the intolerance of such overlaps, but differs in that the inputs are given as pairs of $p$ and $s$ rather than pair of sets or lists $P$ and $S$. Naively solving the pair-of-sets problem using algorithms for the pairs of $p$ and $s$ problem would take $|P|\times|S|$ queries, as we need one query for each pair of $p\in P$ and $s\in S$. This is clearly inefficient for large $|P|$ and $|S|$, and hence the necessity remains for our proposed algorithms.

\section{Conclusion and future work}
In this paper, we proposed online algorithms for counting and reporting all distinct substrings
of an online text $T$ that has some $p\in P$ as a prefix, some $s\in S$ as a suffix, and whose length is within the interval $[k_1..k_2]$,
where $P$ and $S$ are static sequences of strings given as input for preprocessing,
while positive integers $k_1, k_2$ and the characters of $T$ are given as query.
Our algorithms take $O((\Vert P\Vert + \Vert S\Vert )\log \sigma)$ preprocessing time for general ordered alphabets,
which is reduced to $O(\Vert P\Vert + \Vert S\Vert)$ time for integer alphabets of size polynomial w.r.t. $\Vert P\Vert + \Vert S\Vert$.
The computation up to the $i$-th character of $T$ takes $O(|T_i|\log \sigma)$ cumulative time for the counting problem, and $O(|T_i|\log \sigma + |\ans_i|)$ cumulative time for the reporting problem.
Furthermore, we have shown that it has possible applications in traffic classification, by showing that all of the application signatures in ~\cite{Sen2004} can be represented as input sets of our proposed problems.

A few problems remain to be considered as future work:
\begin{itemize}
\item As the discussion in Section \ref{SecSignatures} implies,
solving the problem where the prefix and suffix strings are not allowed to overlap,
i.e. substrings are in the form of $p\Sigma^k s$, where $k\in [k_1..k_2], p\in P, s\in S$,
while retaining the distinctness condition as well as that the input sets be given as pairs of lists $P$ and $S$,
can be useful in case we have a signature implemented with $P, S$ such that there does exist a string $w$ such that $w$ has $p\in P$ as prefix, $s\in S$ as suffix and $|w|<|p|+|s|$, and we want to exclude such $w$ from matches.
Is it possible to devise an algorithm that solve this problem efficiently?

\item In practice, how do the running times of our algorithms compare to the signature implementations used in \cite{Sen2004}?
%\item In practice, when using the input sets specified in Section \ref{SecSignatures}, how do the running times of our algorithms compare to the signature implementations used in \cite{Sen2004}?

%\item Some community rules in Snort \cite{Snort} are in the form of prefix, suffix and length conditions. Can we express several community rules as an input set of our proposed problems, such that we can simulate the checking of multiple Snort rules with one instance of our proposed problems to speedup checking the rules?
\end{itemize}

%general alphabet -> general ordered alphabets
  
%Before formatting:
%  For Gnutella signatures, let
%$P=\text{\{User-Agent:,UserAgent:,Server:\}}$, and $S$ be the set of names, i.e. $S=$
%  \{LimeWire, BearShare, Gnucleus, MorpheusOS, XoloX, MorpheusPE, gtkgnutella, Acquisition, Mutella-0.4.1, MyNapster, Mutella0.4.1, Mutella-0.4, Qtella, AquaLime, NapShare,
%  Comeback, Go, PHEX, SwapNut, Mutella-0.4.0, Shareaza,
%  Mutella-0.3.9b, Morpheus, FreeWire, Openext, Mutella-0.3.3,
%  Phex\}.
%  For eDonkey signatures, let $P$ consist solely of 0xe3 in hex, $S$ consist solely of the packet length, and $k_1=k_2=$5 bytes.
%  For DirectConnect, let $P=$\{\$MyNick, \$Lock, \$Key, \$Direction, \$GetListLen, \$ListLen, \$MaxedOut, \$Error, \$Send, \$Get, \$FileLength, \$Canceled, \$HubName, \$ValidateNick, \$ValidateDenide, \$GetPass, \$MyPass, \$BadPass, \$Version, \$Hello, \$LogedIn, \$MyINFO, \$GetINFO, \$GetNickList, \$NickList, \$OpList, \$To, \$ConnectToMe, \$MultiConnectToMe, \$RevConnectToMe, \$Search, \$MultiSearch, \$SR, \$Kick, \$OpForceMove, \%$ForceMove, \$Quit\} and $S=$\{|\}.
%  For BitTorrent, let $P=S=$\{19BitTorrent protocol\} and $k_1=k_2=\text{20 bytes}$.
%  For Kazaa, let $P=\text{\{GET,HTTP\}}$ and $S=\text{\{X-Kazaa\}}$.

\section*{Acknowledgements}

This work was supported by JSPS KAKENHI Grant Numbers JP20H04141 (HB) and
JP22H03551 (SI), and by JST PRESTO Grant Number JPMJPR1922 (SI). 
  
%\clearpage
  
\bibliographystyle{splncs04}
\bibliography{ReferencesList}

\clearpage

\appendix
\section{Appendix}
\label{AppendixInputSet}
  Below, we show the input sets that match the each of the application signatures described in~\cite{Sen2004}.

\begin{table}

\caption{Input sets corresponding to application signatures}

\begin{tabular}{ | m{2.5cm} | m{4.5cm}| m{4.5cm} | m{0.8cm} | m{0.8cm} | } 

  \hline
  %Application & List of $P$ elements & List of $S$ elements & $[k_1, k_2]$ \\ 
  Application & List of $P$ elements & List of $S$ elements & $k_1$ & $k_2$ \\ 
  \hline

  Gnutella & 

\raggedright 
$\mtt{User-Agent:},$ $\mtt{UserAgent:},$ $\mtt{Server:}$ & 
\raggedright
$\mtt{LimeWire}, $  
$  \mtt{BearShare}, $  
$  \mtt{Gnucleus}, $  
$  \mtt{MorpheusOS}, $  
$  \mtt{XoloX}, $  
$  \mtt{MorpheusPE}, $  
$  \mtt{gtkgnutella}, $  
$  \mtt{Acquisition}, $  
$  \mtt{Mutella-0.4.1}, $  
$  \mtt{MyNapster}, $  
$  \mtt{Mutella0.4.1}, $  
$  \mtt{Mutella-0.4}, $  
$  \mtt{Qtella}, $  
$  \mtt{AquaLime}, $  
$  \mtt{NapShare}, $  
$  \mtt{Comeback}, $  
$  \mtt{Go}, $  
$  \mtt{PHEX}, $  
$  \mtt{SwapNut}, $  
$  \mtt{Mutella-0.4.0}, $  
$  \mtt{Shareaza}, $  
$  \mtt{Mutella-0.3.9b}, $  
$  \mtt{Morpheus}, $  
$  \mtt{FreeWire}, $  
$  \mtt{Openext}, $  
$  \mtt{Mutella-0.3.3}, $  
$  \mtt{Phex}$   &
 $1$ & $\infty$ \\
 
\hline
 eDonkey &
$ \mtt{0xe3}$ (in hex) &
(the packet length)&
\multicolumn{2}{m{1.6cm}|}{5-byte long}\\
%$[$5 bytes, 5 bytes$]$\\

%SPLIT if needed?
%\hline
%\end{tabular}
%\begin{tabular}{ | m{2.5cm} | m{4.5cm}| m{4.5cm} | m{0.8cm} | m{0.8cm} | } 

\hline
DirectConnect &
\raggedright
$ \mtt{\$ MyNick},$ 
$ \mtt{\$ Lock}, $ 
$ \mtt{\$ Key}, $ 
$ \mtt{\$ Direction}, $ 
$ \mtt{\$ GetListLen}, $ 
$ \mtt{\$ ListLen}, $ 
$ \mtt{\$ MaxedOut}, $ 
$ \mtt{\$ Error}, $ 
$ \mtt{\$ Send}, $ 
$ \mtt{\$ Get}, $ 
$ \mtt{\$ FileLength}, $ 
$ \mtt{\$ Canceled}, $ 
$ \mtt{\$ HubName}, $ 
$ \mtt{\$ ValidateNick}, $ 
$ \mtt{\$ ValidateDenide}, $ 
$ \mtt{\$ GetPass}, $ 
$ \mtt{\$ MyPass}, $ 
$ \mtt{\$ BadPass}, $ 
$ \mtt{\$ Version}, $ 
$ \mtt{\$ Hello}, $ 
$ \mtt{\$ LogedIn}, $ 
$ \mtt{\$ MyINFO}, $ 
$ \mtt{\$ GetINFO}, $ 
$ \mtt{\$ GetNickList}, $ 
$ \mtt{\$ NickList}, $ 
$ \mtt{\$ OpList}, $ 
$ \mtt{\$ To}, $ 
$ \mtt{\$ ConnectToMe}, $ 
$ \mtt{\$ MultiConnectToMe}, $ 
$ \mtt{\$ RevConnectToMe}, $ 
$ \mtt{\$ Search}, $ 
$ \mtt{\$ MultiSearch}, $ 
$ \mtt{\$ SR}, $ 
$ \mtt{\$ Kick}, $ 
$ \mtt{\$ OpForceMove}, $ 
$ \mtt{\$ ForceMove}, $ 
$ \mtt{\$ Quit}$ &
$\mtt{|}$ &             %S elements
 $1$ & $\infty$ \\

\hline
BitTorrent &
  %\multicolumn{2}{c}{ the 20-byte string where the first byte is 19 (0x13) and the next 19 bytes are the string $\mtt{19BitTorrent\; \, protocol}$ } &
%  \multicolumn{2}{l|}{ the byte 19 (0x13) followed by $\mtt{19BitTorrent\; \, protocol}$ } &

  \multicolumn{2}{m{9cm}|}{ \raggedright the 20-byte string where the first byte is 19 (0x13) and the next 19 bytes are the string $\mtt{19BitTorrent\; \, protocol}$ } &
\multicolumn{2}{m{1.6cm}|}{20-byte long}\\
  %For Kazaa, let $P=(\mtt{GET},\mtt{HTTP})$ and $S=(\mtt{X-Kazaa})$.
  
  \hline
Kazaa &
$\mtt{GET}$, $\mtt{HTTP}$ &
$\mtt{X-Kazaa}$ &
 $1$ & $\infty$ \\
  \hline
\end{tabular}

\end{table}

\end{document}